\documentclass[sigplan,table,dvipsnames,table,10pt,preprint]{acmart}
\usepackage[utf8]{inputenc}
\usepackage{xcolor}
\usepackage{fixltx2e}
\usepackage{dblfloatfix}
\usepackage{amsmath}
\usepackage{amssymb}
\usepackage{amsfonts}
\usepackage{graphicx}
\usepackage{epsfig}
\usepackage{epstopdf}
\usepackage{multirow}
\usepackage{tabularx}
\usepackage{flushend}
\usepackage{array}
\usepackage{booktabs}
\usepackage{lscape}
\usepackage{nicefrac}
\usepackage{lipsum}
\usepackage{tikz}
\usepackage{balance}
\usetikzlibrary{arrows}
\usepackage{subcaption}
\usepackage{rotating}
\usepackage{soul}

\usetikzlibrary{patterns}
\usepackage{blkarray}
\usepackage{standalone}
\usepackage{pgfplots}

\usepackage{xspace}
\usepackage{bm}
\usepackage{url}
\usepackage{paralist}

\hypersetup{
  colorlinks,
  linkcolor={green!50!black},
  citecolor={red!70!black},
  urlcolor={blue!70!black}
}

\copyrightyear{2019} 
\acmYear{2019} 
\acmConference[Middleware '19]{Middleware '19: Middleware '19: 20th International Middleware Conference}{December 8--13, 2019}{Davis, CA, USA}
\acmBooktitle{Middleware '19: Middleware '19: 20th International Middleware Conference, December 8--13, 2019, Davis, CA, USA}
\acmPrice{15.00}
\acmDOI{10.1145/3361525.3361547}
\acmISBN{978-1-4503-7009-7/19/12}

\def \DA {\textsc{DiAS}\xspace}

\newcommand{\bs}{\boldsymbol}

\newcommand{\cR}{{\cal R}}

\newcommand{\squishlist}{
 \begin{list}{$\bullet$}
  { \setlength{\itemsep}{0pt}
     \setlength{\parsep}{3pt}
     \setlength{\topsep}{3pt}
     \setlength{\partopsep}{0pt}
     \setlength{\leftmargin}{1.5em}
     \setlength{\labelwidth}{1em}
     \setlength{\labelsep}{0.5em} } }

\newcommand{\squishend}{
  \end{list}  }

\graphicspath{{./}}

\begin{document}

\settopmatter{printacmref=false} % Removes citation information below abstract
\renewcommand\footnotetextcopyrightpermission[1]{} % removes footnote with conference information in first column
\pagestyle{plain} % removes running headers

\title[Differential Approximation and Sprinting]{Differential Approximation and Sprinting for Multi-Priority Big Data Engines}
%\title[Optimizing Multi-Priority Big Data Engines\\via Differential Approximation and Sprinting]{Optimizing Multi-Priority Big Data Engines via Differential Approximation and Sprinting}
%% Author information
%% Contents and number of authors suppressed with 'anonymous'.
%% Each author should be introduced by \author, followed by
%% \authornote (optional), \orcid (optional), \affiliation, and
%% \email.
%% An author may have multiple affiliations and/or emails; repeat the
%% appropriate command.
%% Many elements are not rendered, but should be provided for metadata
%% extraction tools.

%% Author with single affiliation.
\author{Robert Birke}
\affiliation{
  \institution{ABB Research}
  \city{Baden-Dättwil}
  \country{Switzerland}
}
\email{robert.birke@ch.abb.com} 

\author{Isabelly Rocha}
\affiliation{
  \institution{University of Neuchâtel}
  \city{Neuchâtel}
  \country{Switzerland}
}
\email{isabelly.rocha@unine.ch}   

\author{Juan Perez}
\affiliation{
  \institution{Universidad del Rosario}
  \city{Bogotá}
  \country{Colombia}
}
\email{juanferna.perez@urosario.edu.co}          

\author{Valerio Schiavoni}
\affiliation{
  \institution{University of Neuchâtel}
  \city{Neuchâtel}
  \country{Switzerland}
}
\email{valerio.schiavoni@unine.ch} 

\author{Pascal Felber}
\affiliation{
  \institution{University of Neuchâtel}
  \city{Neuchâtel}
  \country{Switzerland}
}
\email{pascal.felber@unine.ch} 

\author{Lydia Y. Chen}
\affiliation{
  \institution{TU Delft}
  \city{Delft}
  \country{Netherlands}
}
\email{y.chen-10@tudelft.nl}

%%
%% By default, the full list of authors will be used in the page
%% headers. Often, this list is too long, and will overlap
%% other information printed in the page headers. This command allows
%% the author to define a more concise list
%% of authors' names for this purpose.
\renewcommand{\shortauthors}{R. Birke, I. Rocha, J. Perez, V. Schiavoni, P. Felber, and L. Y. Chen}

\begin{abstract}
	Today's big data clusters based on the MapReduce paradigm are capable of executing analysis jobs with multiple priorities, providing differential latency guarantees.
	Traces from production systems show that the latency advantage of high-priority jobs comes at the cost of severe latency degradation of low-priority jobs as well as daunting resource waste caused by repetitive eviction and re-execution of low-priority jobs.
	We advocate a new resource management design that exploits the idea of differential approximation and sprinting. %, \emph{i.e.}, providing different levels of accuracy and computation frequency to jobs with different priority levels. 
	The unique combination of approximation and sprinting avoids the eviction of low-priority jobs and its consequent latency degradation and resource waste.
	%Instead of evicting low-priority jobs, differential approximation deflates their workload, abiding by their minimum accuracy requirements, thereby making resources available for high-priority jobs.
	%Differential approximation is thus able to reduce resource waste and improve the latency of low-priority jobs by avoiding evictions and re-executions.
	To this end, we designed, implemented and evaluated \DA, an extension of the Spark processing engine to support deflate jobs by dropping tasks and to sprint jobs. % at a higher frequency.
	%\DA leverages a set of stochastic models based on matrix analytic methods.
	%The core components of \DA are priority buffers and a \rb{task}{job} deflator, which not only deflates the workload but is also capable of modifying this choice based on the latency model it implements.
	%We evaluate \DA on text retrieval and graph mining applications.
	Our experiments on scenarios with two and three priority classes indicate that \DA achieves up to 90\% and 60\% latency reduction for low- and high-priority jobs, respectively. 
	%Compared to a baseline Spark with preemptive eviction of low-priority jobs, 
	\DA not only eliminates resource waste but also (surprisingly) lowers energy consumption up to 30\% at  only a marginal accuracy loss for low-priority jobs. 
\end{abstract}

%%
%% The code below is generated by the tool at http://dl.acm.org/ccs.cfm.
%% Please copy and paste the code instead of the example below.
%%
%\begin{CCSXML}
%	<ccs2012>
%	<concept>
%	<concept_id>10010520.10010553.10010562</concept_id>
%	<concept_desc>Computer systems organization~Embedded systems</concept_desc>
%	<concept_significance>500</concept_significance>
%	</ccs2012>
%\end{CCSXML}

%\ccsdesc[500]{Computer systems organization~Embedded systems}

%%
%% Keywords. The author(s) should pick words that accurately describe
%% the work being presented. Separate the keywords with commas.
%\keywords{datasets, neural networks, gaze detection, text tagging}

\maketitle

%!TEX root = middleware_19.tex
\section{Introduction}\label{sec:introduction}

Big data production systems, \emph{e.g.}, Google~\cite{Clusterdata:Wilkes2011} and Facebook~\cite{Chen:VLDB12:FacebookClaudera}, implement priority scheduling to process job streams with different characteristics and latency requirements.
Analysis jobs, \emph{e.g.}, hive queries~\cite{ganapathi2010statistics} and text mining, of different priorities arrive in streams and are executed as parallel jobs with varying numbers of map and reduce tasks.  %that last for different duration.
Trace studies~\cite{Cavdar:greenmetrics14:priority} show that high-priority jobs are promptly served with little queueing time, while low-priority jobs suffer from repetitive evictions causing significant resource waste.
This can be attributed to the practice of preemptive priority scheduling~\cite{Schwarzkopf:Eurosys13:omega}, where high-priority jobs are given %the highest precedence with
the ability to preempt lower-priority jobs in execution.
The average latency slowdown of low-priority jobs~\cite{Rosa:DSN15:failure}, \emph{i.e.}, the end-to-end response time divided by the execution time excluding eviction, can be 3$\times$ higher than for high-priority jobs.
All in all, priority-enabled big data systems preserve the performance advantage of high-priority jobs at the cost of resource efficiency and performance of low-priority jobs.

It is extremely challenging to optimize the performance of big data %jobs in priority systems
engines with priority scheduling as performance conflicts arise across disparate job priorities as well as across performance targets, %measurements,
\emph{i.e.}, latency \emph{vs.} resource efficiency.
Meeting the latency targets in priority systems is a long standing challenge from both system~\cite{Schwarzkopf:Eurosys13:omega, Cho:SoCC13:Natjam} and modelling~\cite{Mor:Questa05:priority,Horvath:ejor15:priority} perspectives due to the complex dynamics and performance requirements across diverse priority classes.
This is partly because the processing order of low-priority jobs highly depends on the high-priority jobs, especially when low-priority jobs are evicted during periods of resource shortage.
In additional to the inter-job dependency, big data jobs themselves have complex execution dynamics across their parallel tasks and synchronization stages. 
% In addition, for big data jobs it is also key to consider that they are composed of parallel tasks, such that the job latency %/accuracy
%highly depends not only on the inter-job dynamics but also on the interdependency across tasks, e.g., the latency of a job is determined by the slowest task. To make things more complicated, a typical big data job progresses in stages with one or more synchronization periods during its lifetime.

Existing systems address the latency issue of big data engines mainly from the perspective of single job type, \emph{i.e.}, one single priority. 
On the one hand, approximation-enabled processing engines, \emph{e.g.}, BlinkDB~\cite{Agarwal:Eurosys:BlinkDB} and ApproxHadoop~\cite{Goiri:ASPLOS15:APPROXHADOOP}, reduce the execution times by processing a fraction of input data. 
The performance advantage comes at the cost of accuracy losses. 
On the other hand, hardware features are increasingly being exploited to accelerate the executions of jobs. 
For example, Pupil~\cite{zhang2016pupil} and Sprinting Game~\cite{Fan:ASPLOS16:SprintGame} temporarily sprint the CPU frequency during the slow execution phases of jobs. 
However, these engines do not readily apply to multi-priority scenarios, answering the performance  and resource tradeoff among different~priorities.

We advocate to differentially approximate and sprint CPU frequency for jobs of different priorities, termed \emph{differential approximation and sprinting} (\DA), to replace preemptive eviction in priority scheduling. 
\DA improves the latency for all priorities and eliminates resource waste from re-executing the evicted low-priority jobs. 
To achieve this, \DA reduces a fraction of data load for low-priority jobs and temporarily increase the CPU frequency for high-priority~jobs. 

The differential approximation adopts a controllable approximation level that discriminates among priority classes by dropping different fractions of data.
It gives better latencies for low-priority jobs at the cost of their accuracy loss and minor latency increase for high-priority jobs, depending on the levels of input data dropping.
The differential sprinting then adjusts the frequency levels such that the high-priority jobs can be accelerated after temporarily waiting behind the low-priority approximate jobs.

We design, implement and evaluate \DA on top of \sloppy{Spark~\cite{zaharia2016apache}}. %taking advantage of its native support for approximation mechanisms.
The \DA extension module is composed of $N$ priority buffers,  and one deflator %\vs{define this}
 that assigns approximation and sprinting levels for each priority.
To determine the dropping and frequency levels, we derive a set of stochastic models that can predict average response times of \DA jobs.
The models, based on matrix-analytic methods and parameterized via simple linear regressions, can effectively guide the choice of approximation levels for each priority. 

We evaluate \DA using benchmarks that process the contents of the \emph{stackexchange}~\cite{stackexchange} network of question-and-answer websites as well as the Google web graph~\cite{google}, with two and three job priority levels.  %analyzing web contents.
To demonstrate the robustness of \DA, we consider various workload profiles, \emph{i.e.}, priority ratios, job sizes, and system loads.
Evaluation results show that \DA achieves remarkable reductions not only on the mean and tail latency of low-priority jobs, but also on the \emph{tail latency of high-priority jobs}.
With \DA we achieve up to 90\% and 60\% reduction in the mean/tail latency for low-priority and high priority jobs, respectively, a preemptive priority system without approximation and sprinting.  Moreover, the promising performance gain of \DA comes with a noticeable energy reduction energy, i.e., up to 30\% , even after spending extra power to sprint the high priority~jobs.

Our contributions are multi-fold. First, we put forward a first of its kind design for differential approximation and sprinting that preserves the latency advantage of high-priority jobs and reverts the latency disadvantage of low-priority jobs for both mean and tail latencies. Second, we derive bottom-up stochastic models that capture the dynamics of big data jobs (at both the task and the stage levels) that implement different approximation and sprinting levels. Third, we implemented \DA on top of Spark, the state-of-the-art big data processing engine, by building a model-based job deflator and augmenting Spark with the approximation capability of dropping tasks.  Last, \DA agilely combines multiple knobs (task dropping, sprinting, and scheduling) to achieve significant latency and energy reduction from the state of the~art.
%!TEX root = middleware_19.tex
\section{Motivation and Background}\label{sec:motivation}

Here, we first use traces collected from production systems to motivate the performance pitfall of preemptive priority scheduling, i.e., resource waste from evicting low priority jobs.
We then discuss the background of priority scheduling, big data processing engines to highlight their complexity, and computational sprinting.

\subsection{Resource Waste in Production Systems}
A number of field studies~\cite{Rosa:DSN15:failure,Rosa:Perv15:Eviction, Clusterdata:Wilkes2011, Schwarzkopf:Eurosys13:omega} from publicly available big data cluster traces show that priority scheduling is widely adopted in production big data systems.
Workloads are defined in the unit of jobs that are in turn composed of multiple tasks. Jobs are divided into multiple classes, each of which is assigned with a priority level. For example, Google clusters employ 12 priority levels~\cite{Schwarzkopf:Eurosys13:omega}. %, ranging from priority 0 to 11.
Performance of high-priority jobs is typically enforced at the cost of low-priority jobs.
Earlier studies~\cite{Rosa:Perv15:Eviction} show that jobs with the lowest priority (priority 0) are repetitively evicted by the scheduler, due to the arrival of high priority workloads.
The unfortunate consequences of eviction in production systems are two-fold: (i) high amount of resources, \emph{i.e.}, 25\% machine and 30\% CPU time, are spent/wasted on subsequent evictions; (ii) significant latency degradation for low priority jobs,  \emph{e.g.}, the slowdown of priority-0 jobs compared to the case of no eviction is 3 times higher than that of priority-6 jobs.
Differential approximation eliminates this resource waste as low-priority jobs are never evicted. Instead, low priority jobs are processed approximately to provide short response times and to allow high-priority jobs to quickly gain access to processing slots.

\subsection{Priority Scheduling}

\begin{figure}[t]
  \centering
  %\begin{tabular}{c}
   %\includegraphics[scale=0.8]{figs/queues.pdf}\\[12pt]
   \includegraphics[scale=1.0]{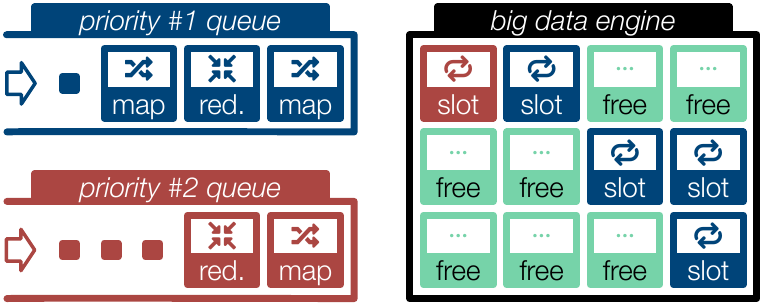}
   %        \end{tabular}
    \caption{Schematics of priority scheduling for big data jobs.}
    \label{fig:priority}
\end{figure}

It is of paramount importance to optimize the priority scheduler, especially when encountering big data workloads with strong inter- and intra-priority dynamicity~\cite{Chen:VLDB12:FacebookClaudera,Schwarzkopf:Eurosys13:omega}.
Priority schedulers typically separate jobs by their priority levels and keep them in separate queues.
Then, they determine when and which job to process next %at the analysis framework%, such as Spark, and allocate slots to jobs. 
Figure~\ref{fig:priority} depicts the schematics of a big data cluster with multiple slots serving high- and low-priority jobs in two queues.
Jobs arrive with a number of parallel tasks, following MapReduce~\cite{dean2008mapreduce} programming paradigm and having time-varying arrival rates.
%MapReduce is a parallel programming paradigm to process data at scale, and Spark~\cite{Spark} is a popular open source implementation of the same with additional support for fast iterative computations and fault tolerance.
Jobs with the same priority class stay in the same queue and are typically served in a first-come-first-served (FCFS) manner.
Across priorities, jobs with a higher priority have precedence over any job with a lower priority.
Upon arrival of a high priority job, the scheduler ensures that it is served quickly by either evicting any lower priority jobs currently being processed,  or by letting the current lower priority jobs to finish and immediately start the incoming job.
%before allowing the new  %a pre-configured delay.
The former is called preemptive, whereas the latter is termed non-preemptive priority scheduling.
As observed in~\cite{Rosa:Perv15:Eviction}, the Google production systems employ preemptive priority scheduling, causing significant resource waste due to evictions.
After being evicted, low-priority jobs return to the head of the queue and wait for new scheduling opportunities, \emph{i.e.}, until no high-priority job is queued or processed.
These evictions are completely avoided by differential approximation as it employs non-preemptive scheduling, with the consequent resource savings.

\subsection{Computational Sprinting}

Computational sprinting allows for bursts of peak performance under a sprinting budget.
The sprinting budget can stem from thermal~\cite{DBLP:journals/micro/RotemNAWR12}, power~\cite{Fan:ASPLOS16:SprintGame} or provisioning~\cite{AWS:burstable} constraints.
Several sprinting mechanisms exist, from modifying the CPU performance via dynamic voltage and frequency scaling (DVFS)~\cite{DBLP:journals/micro/RotemNAWR12} to tuning the job parallelism~\cite{DBLP:conf/asplos/HaqueEHEBM15}. 
The common aim is to temporarily accelerate the execution of a job.  
Each sprinting mechanism is controlled via a corresponding sprinting policy which determines what and when to sprint.
Time-based policies levering timeouts to control the sprinting are rather common~\cite{DBLP:conf/hpca/HsuZLMWMTD15,Morris:Eurosys18:sprinting}.
%A common choice are Time-based policies where timeouts control the sprinting~\cite{DBLP:conf/hpca/HsuZLMWMTD15,Morris:Eurosys18:sprinting}.

\subsection{Processing Engines and MapReduce Jobs}

MapReduce is a parallel programming paradigm to process data at scale. 
Spark~\cite{Spark} is a popular open source implementation of this paradigm with additional support for fast iterative computations and fault tolerance mechanisms.
%%In the remainder of this paper, we use a Spark job to illustrate the basic processing flow of analysis jobs.
A typical MapReduce job processes input data and returns analysis results via parallel tasks executed in multiple map and reduce stages.
The input data are organized as blocks stored in a file system, such as the Hadoop File System (HDFS)~\cite{shvachko2010hadoop}, which splits data across the servers in the cluster.
During a map stage, map tasks are spawned to process one input block each.  
Intermediate results are stored as key-value pairs. % in the file system.
Afterwards, reduce tasks access and aggregate the intermediate key-value pairs for the final result. 
A job can comprise multiple such map and reduce tasks/stages.
Specific policies supported by cluster job scheduler exist to allow one or multiple concurrent jobs in the engine, respectively for FCFS and weighted fair sharing.

\DA provides a turnkey solution compatible with the plethora of existing scheduling frameworks and job schedulers (\emph{e.g.}, Hadoop and YARN~\cite{Vavilapalli:2013:AHY:2523616.2523633}) that support various allocation policies and resources across analysis frameworks.
In essence, \DA achieves this by altering the job sizes to fulfill the latency and precision constraints simultaneously.
We detail this approach next.
%%Our focus here is to improve the job scheduler that resides in the analysis framework.
%The proposed solution, \DA,  described in the next section is compatible with different scheduling frameworks and job schedulers, with the objective of altering job sizes to fulfill the latency and precision constraints simultaneously.
%!TEX root = middleware_19.tex

\section{Differential Approximation and Sprinting}\label{sec:Solution}

\begin{figure}
  \centering
   \includegraphics[scale=1.0]{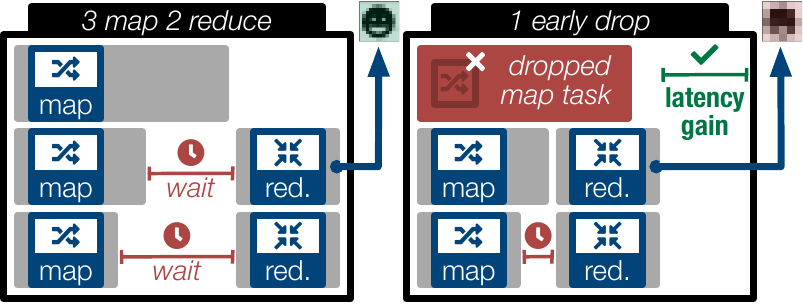}
    \caption{Approximation by task dropping: an example on a job of 3 maps and 2 reduce tasks.} \label{fig:Taskdrop}
\end{figure}

Motivated by the importance and complexity of tuning the performance of priority-enabled systems, we propose the idea of differential approximation and sprinting across different priority levels as a means to \begin{inparaenum}[\it (i)]
 \item reshape the workload demands of jobs in each priority level;
 \item implicitly provision more resources to higher-priority jobs
 \item speed up the execution of lower-priority jobs by deflating their processing load, \emph{i.e.}, number of tasks; 
 \item provide consistent performance guarantees on high-priority jobs, and 
 \item minimize resource waste.
\end{inparaenum}
The core goal of differential approximation and sprinting is to decide the approximation level (task dropping ratio), denoted by $ 0\le \theta_k \le 1$, and sprinting timeout, denoted by $T_k$ to be applied on arriving jobs given their priority class $k$, their tolerance to accuracy degradation, and the available sprinting budget. % tolerances in terms of relative errors.
The expected outcome of differential approximation in a scenario of two job priorities, \emph{i.e.}, high \emph{vs.}~low, is to minimize the resource waste and average/tail latencies of high/low priority jobs, while maintaining the relative error of low priority jobs within a given bound and fully use the available sprinting budget. 

In contrast to preemptive schedulers, we alter the resource demand of lower- and higher-priority jobs, instead of evicting lower-priority jobs upon the arrival of higher-priority jobs.
\DA is our implementation of this design. 
It plugs into existing big data processing engines to support differential approximation, computational sprinting and workload deflation by means of dropping tasks. 
%To such an end, we develop \DA, a prototype module for big data processing engines supporting differential approximation, computational sprinting and workload deflation by means of dropping tasks. 
%Prior to presenting the architecture, we first describe how to achieve approximation analysis for a single job at the processing engine and describe its accuracy impact.

\subsection{Approximate Big Data Jobs}

The aim of approximate computing in big data processing~\cite{Agarwal:Eurosys:BlinkDB,Goiri:ASPLOS15:APPROXHADOOP} is to solve the performance conundrum between latency and accuracy requirements of analysis jobs.
Instead of processing all the input data, only a subset of data is chosen to be processed to lower the overall computation demand and reduce latency. 
Existing systems (\emph{e.g.}, ApproxHadoop~\cite{Goiri:ASPLOS15:APPROXHADOOP}) put a significant engineering effort to enable dropping (map) tasks and their assigned input data prior or during execution. %, which are usually engine-dependent.
%To illustrate the task dropping strategy to attain certain approximation level, let us take an example of a simple job that has 3 map and 2 reduce tasks as shown in Fig.~\ref{fig:Taskdrop}. 
Figure~\ref{fig:Taskdrop} illustrates this task dropping strategy on a simple job with 3 map and 2 reduce tasks, with the goal of attaining a given approximation level.
%There are two possible choices to process the map tasks, namely (i) process them all, or (ii) randomly drop some tasks before their execution. 
In this example, we randomly choose one map task and drop it before its execution. 
Task dropping saves the overhead of fetching data and avoids the execution of the dropped tasks. 
Nevertheless, while it reduces the computational demand of jobs, it unavoidably degrades the analysis accuracy.
%While dropping tasks reduces the %total% amount of
%computational demand of jobs, %there is no guarantee on reducing the job processing time.% for the following reason.
%Production traces~\cite{Chen:VLDB12:FacebookClaudera} show that the number of (map) tasks is usually higher than the amount of parallelism on a server, thus execution is done in multiple waves and job processing time is determined by the slowest task that completes in all waves.
%Nonetheless, processing a subset of data accelerates the job time, but
%it unavoidably degrades the analysis accuracy. % of the analysis.
This precision loss depends on both the analysis performed and the data, %dependent 
and can be estimated offline as shown in Section~\ref{sec:expsetup}.

\begin{figure}
  \centering
  \includegraphics[scale=1.0]{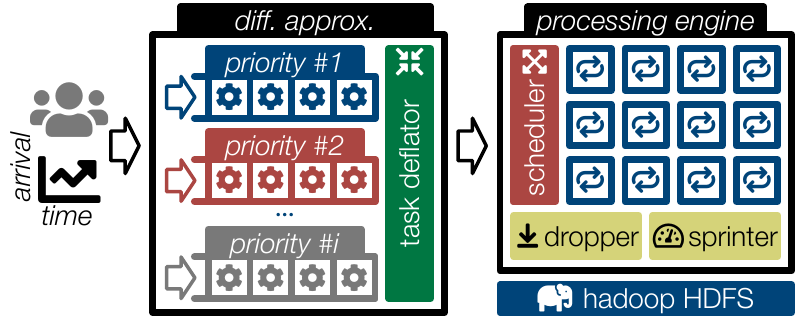}
  \caption{Schematics of differential approximation.}\label{fig:DA}
\end{figure}

\subsection{Architecture of the \DA prototype}
Figure~\ref{fig:DA} depicts the architecture of \DA.
The key components of \DA are:  \begin{inparaenum}[\it (1)] \item a set of job buffers for each priority, 
indexed by $k \in \{0, \dots, K\}$, \item the task deflator that determines the approximation level $\theta_k$ and sprinting timeout $T_k$ for each priority $k$, and \item the sprinter which temporarily sprints jobs\end{inparaenum}.
Higher values of $k$ indicate higher priority.
The deflator has two main functionalities: (i) to determine the approximation level and sprinting timeout based on empirical/stochastic models as well as on performance and budget thresholds, and (ii) to dispatch jobs from the priority buffers into the processing engine. 
We first describe how the deflator dispatches and evicts jobs 
%in the subsequent sections
and derive the analytical models in Section~\ref{sec:model}.

Upon arrival, jobs are immediately dispatched to the corresponding buffer according to their priorities.
Jobs queued at each buffer are processed in a FCFS manner.
The task deflator selects the job at the head of the highest non-empty priority buffer, say $k$. 
This priority-$k$ job with its corresponding approximation level, $\theta_k$, is sent to the processing engine, which splits the job into multiple tasks on the cluster. %, and a timer with timeout $T_k$ is started. % using different task scheduling policies. 
To avoid potential resource waste caused by eviction using preemption, the execution across priority buffers of \DA is non preemptive. 
\DA only dispatches jobs from the head of the buffer to the processing engine when the previous job completes, independent of the priority of newly arrived jobs.
Moreover, \DA assumes that the processing engine is able to drop tasks to achieve the target ratio $\theta_k$. 
We note that as \DA aims to reshape the job workloads prior to entering the processing engines, the design of \DA is general and compatible with different processing engines.

For our baseline results with preemption, \DA also provides the ability to evict jobs from the processing engine. 
In this case, as soon as any job arrives with a higher priority than the job currently being processed, the job in the engine is evicted back to the head of its buffer and the arriving job is immediately sent in for execution.

In addition, \DA can leverage computational sprinting to further counter the effects on higher priority jobs stemming from not preempting lower priority ones.
If sprinting is enabled, the deflator in parallel to dispatching the job communicates to the sprinter the sprinting timeout $T_k$ to use.
Once the timeout elapses, the sprinter will temporarily accelerate the execution of the job.

%Evicted jobs can either resume from where they have been evicted or start from scratch, contingent on the availability of check-pointing mechanism provided by the processing engine.
%When the overhead of check-pointing is high, the benefit of resuming might be overturned and non-resuming can be a preferred option.
\subsection{Implementation}
We implement the \DA prototype in the Go programming language and use Spark as big data processing engine.
In addition to priority buffers and the deflator, we also implemented a workload generator and augment Spark with the capability to drop tasks.
%task dropping on Spark and a workload generator. %\rb{that can generate jobs in statistically-distributed inter-arrival times.}{}
To deliver the aforementioned functionalities, \DA is designed to be multi-threaded. %: two threads for deflator and one thread for the generator.

\textbf{ Deflator.}  The deflator consists of one dispatcher and one monitor thread. 
When a job completes or a new job arrives, the dispatcher thread selects which job to run and dispatches it using the {\tt os.exec} library.
It does so by first creating a {\tt cmd} structure and then launches the process with {\tt Start()}. 
When evicting jobs, this thread sends the \texttt{SIGKILL} to the process using {\tt cmd.Process.Kill()}.
The monitor thread surveils the running job, collecting its exit status via {\tt Wait()} and actively relays the completion/eviction of the job to the dispatcher thread using a golang channel.

\textbf{ Sprinter.}
If sprinting is enabled, the sprinter handles a sprinting timer for each dispatched job and tracks the remaining sprinting budget. 
When the timer fires, it uses DVFS to temporarily accelerate the job execution by adjusting the frequency of the CPU on the cluster nodes via the {\tt cpupower} utility. A job is accelerated until either its end or the depletion of the sprinting budget. 
The sprinting budget is replenished over time using a replenishing rate, \emph{e.g.}, 6 sprinting minutes per hour~\cite{AWS:burstable}. 
The timeout is ignored if the job ends sooner. % the timeout, the timeout is ignored. 

\textbf{ Dropper.}
A Spark job typically analyzes a dataset stored as files in HDFS.
% Each Spark job comprises a DAG of stages using Resilient Distributed Datasets (RDD) as input/output.	
Each Spark job is translated into a DAG of operations on Resilient Distributed Datasets (RDD) used as input/output.
The job execution proceeds in stages (\emph{i.e.}, periods of synchronization points). 
Each RDD is made of multiple partitions, the number of which indicates the parallelism achievable by Spark, as each partition can be concurrently executed by only one task.
The size of a job is thus conventionally defined by the number of RDDs and their partitions (equivalently tasks).
%\sloppy{
Each stage relies on the {\tt findMissingPartitions()} function to get the number of partitions to be computed.
%}
To implement task dropping in Spark, %we leverage a virtual function, {\tt findMissingPartitions()}, which can be used to query which partition are yet to be finished in each stage.
we modify {\tt findMissingPartitions()}
% of {\tt SuffleMapStage}
to return only $\lceil n (1-\theta_k) \rceil$ partitions out of $n$ following the specifications of the deflator.

%\RB{I removed the differnce between shufflestage and results stage unless we wnat to explain why dropping results partitions is bad (i.e. aggregate results)}

% The basic workload unit at Spark is a partition of , which can be processed by only one concurrent task.
% Each spark job is translated into a DAG of operations on RDD in stages.
% There are typically of two types of stages, {\tt ShuffleMapStage } and {\tt ResultStage}.
% The virtual function of {\tt findMissingPartitions()} can be used by the Spark scheduler to query which partition are yet to be finished.
% To implement early dropping in Spark, %we leverage a virtual function, {\tt findMissingPartitions()}, which can be used to query which partition are yet to be finished in each stage.
% we modify the {\tt findMissingPartitions()} of {\tt SuffleMapStage} to return only $1-\theta_i$ partitions, specified by the deflator.
%!TEX root = middleware_19.tex
\newcommand{\bN}{\overline N}
\newcommand{\bNm}{\overline N_m}
\newcommand{\bNr}{\overline N_r}
\newcommand{\bmum}{\overline \mu_m}
\newcommand{\bmur}{\overline \mu_r}
\newcommand{\CC}{C}
\newcommand{\bF}{\bs{F}}
\newcommand{\bG}{\bs{G}}
\newcommand{\GG}{G}
\newcommand{\bGs}{\bs{\tilde{G}}}
\newcommand{\bD}{\bs{D}}
\newcommand{\one}{\bs 1}
\newcommand{\0}{\bs{0}}
\newcommand{\I}[1]{\bs{I}_{#1}}
\newcommand{\cO}{\mathcal{O}}
\newcommand{\cM}{\mathcal{M}}
\newcommand{\cS}{\mathcal{S}}
\renewcommand{\cR}{\mathcal{R}}
\newcommand{\cP}{\mathcal{P}}
\section{Modeling \DA}\label{sec:model}
%\subsection{Modeling Priorities}
%We consider Big Data clusters that maintain several queues to provide differentiated service to incoming jobs.
To guide \DA, we analytically derive the response time distribution offered by the cluster to the %of the queuing system that serves
incoming multi-task jobs classified in multiple priorities.
Jobs are classified in $K$ priorities, where a priority-$k$ job has precedence over jobs in priority levels $l<k$, for $1\leq l,k\leq K$.
According to the \DA architecture, jobs are served in FCFS order and each job seizes all the resources in the cluster (or in the partition used by the corresponding engine) to execute.
This can be viewed as a single server queue serving $K$ priority classes.
We thus opt to employ the recent method proposed in~\cite{Horvath:ejor15:priority}, which is capable of obtaining the response time distribution and its moments for a fairly general priority queue with $K$ priority classes under both preemptive and non-preemptive scheduling.
%Further, this method covers preemptive and non-preemptive scheduling, both of which are used by \DA to manage the latency-accuracy tradeoff.

\sloppy{
One key reason to choose~\cite{Horvath:ejor15:priority} as the latency model for \DA is its support for Phase-Type (PH) job processing times~\cite{latouche1999introduction}, which is a class of distributions that can capture fairly general behaviours.}
Further, %This is important because %Also,
PH distributions %are amenable for numerical analysis and
are closed under a number of operations, a feature that we exploit to model the detailed job processing times of concurrent tasks. %In fact, rather than considering a job-level view, where the job service times follow a given distribution, we instead build the job service time distribution from the bottom up, taking a more detailed view at either the task or the wave level.
Instead of using a given distribution to model the job processing time, we resort to a bottom-up approach and build a more detailed view at the task level or wave levels.
For a description of waves and their role in job execution see Section~\ref{sc:wave} below.
%\rb{This is the forst time we speak about waves}
We thus exploit PH distributions to capture details of tasks and waves (\emph{i.e.}, no. of waves =$\lceil{\frac{no. \, tasks}{no. \, slots}} \rceil$) %
%\vs{say what is a wave}\rb{added reference to wave model section} 
within the job processing time and build upon recent results on priority queues with PH components~\cite{Horvath:ejor15:priority}. %for the MMAP[$K$]/PH[$K$]/1 priority queue~\cite{Horvath:ejor15:priority}
%Before detailing how to build these PH distributions at the task and wave levels, we briefly describe background from~\cite{Horvath:ejor15:priority} and the notation.

\emph{Background on the MMAP[$K$]/PH[$K$]/1 priority queue}.
Horv\'ath~\cite{Horvath:ejor15:priority} analytically derived the latency distribution for an MMAP[$K$]/PH[$K$]/1 priority queue, where processing times follow PH distributions, differentiated for the $K$ job classes, and arrivals follow a Marked Markovian Arrival Process (MMAP) with $K$ different streams~\cite{latouche1999introduction}, one for each priority class.
This class of arrival processes can capture fairly general behaviors, including correlations among arrival streams or general inter-arrival times. 
The parameters of an MMAP are $K+1$ $m_a\times m_a$ matrices $(\bD_0, \bD_1,...,\bD_K)$, where $\bD_k$ holds the transition rates for class-$k$ jobs, and $\bD_0$ ensures that the matrix $\bD=\sum_{k=0}^{K}\bD_k$ is the generator of a Markov chain.
The simplest non-trivial example is the marked Poisson arrival process, where $m_a=1$, $\bD_k=\lambda_k$, which is the arrival rate of class-$k$ jobs, and $\bD_0 = -\sum_{k=1}^{K}\lambda_k$.
%\lc{Juan, could we provide reasons why 1 server is a good one though we have C slots?}

\begin{table}[t]
\caption{Summary of notation.}
\label{tb:notation}
\centering
\footnotesize
%\scalebox{0.82}{
\rowcolors{1}{gray!10}{gray!0}
\begin{tabular}{  r | l }
\rowcolor{gray!25}
%\hline
\textbf{Symbol} & \textbf{Definition} \\ \hline
%\multirow{2}{*}{$S$} & $n_w {\geq} 0$
$\CC$ & Number of computing slots\\
$N_m^k$ & Max.~number of map tasks in a priority-$k$ job\\
$N_r^k$ & Max.~number of reduce tasks in a priority-$k$ job\\
\hline
$p_m(t)$ & Prob.~that a priority-$k$ job has $t$ map tasks\\
$p_r(u)$ & Prob.~that a priority-$k$ job has $t$ reduce tasks\\
\hline
$1/\mu_m^k$ & Mean exec.~time for map tasks in a priority-$k$ job\\
$1/\mu_r^k$ & Mean exec.~time for reduce tasks in a priority-$k$ job\\
$1/\mu_o^k$ & Mean setup time for a priority-$k$ job\\
$1/\mu_s^k$ & Mean shuffle time for a priority-$k$ job\\
\hline
$\theta_m^k$ & Approximation ratio for map tasks in a priority-$k$ job\\
$\theta_r^k$ & Approximation ratio for reduce tasks in a priority-$k$ job\\
\hline
$\cO$	& Overhead stage\\
$\cM_t$	& Map stage with $t$ map tasks left to process\\
$\cS$	& Shuffle stage\\
$\cR_u$	& Reduce stage with $u$ map tasks left to process\\
\end{tabular}
%}
\end{table}

\emph{Assumptions and notations on the cluster and approximate/sprinting jobs}.
We assume the cluster, or the allocated partition, is composed of $\CC$ computing slots.
%Low-priority
Priority-$k$ jobs have $n_m^k$ map and $n_r^k$ reduce tasks, both of which are discrete random variables with minimum value 1 and maximum value $N_m^k$ and $N_m^k$, respectively.
%Low-priority jobs have a number of map tasks \rb{$n_m^l$}{$t^l$} and reduce tasks \rb{$n_r^l$}{$u^l$}\rb{, both which are random variables}{}.
On average, the time to execute a map task is $1/\mu_m^k$ and to execute a reduce task is $1/\mu_r^k$.
In addition, the job execution may include an initial setup time that lasts $1/\mu_o^k$ time on average, and an intermediate shuffle stage that requires on average $1/\mu_s^k$ time.
%\rb{check my changes} 
We note that, when sprinting is enabled, the service rates can be approximately captured by the effective sprinting rates as a weighted average of the sprinted and non-sprinted execution times per task and class $k$. Predicting these rates is complex~\cite{Morris:Eurosys18:sprinting}. 
We assume that the effective sprinting rates are provided by an oracle for each class $k$ and timeout value. 
Moreover, as the number of executors available is less than the number of parallel tasks and executors comprise multiple cores, each executor concurrently executes multiple tasks.
Hence, our current approach sprints all available cores at the same time which is beneficial for applications consisting of tasks with equal workloads.
We leave the estimation of effective sprinting rate for complex sprinting policies as future work.
 %integrating the approach from~\cite{Morris:Eurosys18:sprinting} into the deflator as future work.
Table~\ref{tb:notation} summarizes the notation.

%These definitions can be applied to high-priority jobs, which we denote with super-index $h$.}
%\lc{I moved the general description of cluster and job upward, instead of keeping them in task, because they are basic notations used in both models}
%\lc{Juan, could we drop subscript of l and h, and replace by i.}

%In the following, we describe how to model the execution time of jobs that drop a fraction of map or reduce tasks as PH distributions. %We note that the models presented below consider a more general task dropping policy than the prototype of \DA, i.e., both map and reduce tasks can be dropped here.
%\RB{We never said that we drop only map tasks or the reason why we do this. Should we remove this phrase?}

%\begin{figure}[htp]
%  \centering
%  \begin{tabular}{cc}
%    \includegraphics[width=.47\columnwidth]{{{figures/exp-prio-ST-validation-overhead-print_meanRT}}}&
%    \includegraphics[width=.47\columnwidth]{{{figures/exp_RT_valid_exp7_ph10_util80_meanRT}}} \\
%       (a)  Processing Times &  (b)  Response Times
%  \end{tabular}
%  \caption{Validation of job processing and response times for different datasets and priorities. }
%  \label{fig:validation}
%\end{figure}

\subsection{Task-level Model}

For priority-$k$ jobs, we set the task drop ratio to $\theta_m^k$ for map tasks and to $\theta_r^k$ for reduce tasks. The effective number of map and reduce tasks is thus %\rb{
${\bar n_m^k = \lceil n_m^k (1-\theta_m^k)\rceil}$ and ${\bar n_r^k = \lceil n_r^k (1-\theta_r^k)\rceil}$. %}{$\bar t^l = \lceil t^l (1-\theta_m^l)\rceil$ and $\bar u^l = \lceil u^l (1-\theta_r^l)\rceil$}.
Moreover, as the number of tasks is a random variable, we let $p_m^k(t)$ and $p_r^k(u)$ be the probabilities that a priority-$k$ job has %$n_m^k=t${
$t$ map and %\rb{$n_r=u$}{
$u$ reduce tasks, where $1\leq t\leq N_m^k$ and $1\leq u\leq N_r^k$. %, and $N_m^k$ and $N_r^k$ are the maximum number of map and reduce tasks.
%\lc{Juan, could you please check above}
From this point on, we drop the super-index $k$ for clarity, but the definitions apply to all job priorities making use of the appropriate index. %corresponding parameters.

With the above definitions we can extend the model in~\cite{perez2017latency} to incorporate the overhead $\cO$ and shuffle stages $\cS$, as well as to allow for a variable number of tasks.
We thus let the processing phase $i$ %\rb{$J$}{${\cal p}$}
keep track of the job current execution step, where: \begin{inparaenum}[\it (i)]
\item $i=\cO$ indicates the job is in the initial setup (overhead) stage;
\item $i=\cM_t$ indicates that $t$ map tasks remain to be completed, for $1\leq t\leq \bar N_m$;
\item $i=\cS$ indicates the job is in the intermediate shuffle stage;
\item $i=\cR_u$ indicates that $u$ reduce tasks remain to be completed, for $1\leq u \leq \bar N_r$.
\end{inparaenum}
%\RB{I chnaged $J$ to ${\cal p}$}
%\lc{Juan, I changed i to j, please check}\RB{i changed reduce tasks to u}
All jobs start in stage $\cO$ and their evolution is determined by the transition rates from phase $i$ %\rb{$t$}{${\cal p}$}
to the next phase $j$, $f(t,j)$, %}{${\cal p}^+$, $f({\cal p},{\mathcal p}^+)$},
defined as:
%To model the approximate MR execution we describe the job execution time as a phase-type (PH) distribution, where phase $n$ indicates that the job still has $n$ tasks to complete, $1\leq n\leq \bNm+\bNr$.
%A job thus starts processing in phase $\bNm+\bNr$ and moves from phase $n$ to phase $n-1$ with rate $f(n)$ given by
\begin{equation}
f(i, j) = \begin{cases}
				\mu_o p_m(t),		&i = \cO, 		j = \cM_{\bar t}, \\
				\CC \mu_m,			&i = \cM_t, 	j = \cM_{t-1}, t \geq \CC, \\
				t\mu_m,				&i = \cM_t, 	j = \cM_{t-1}, 2\leq t < \CC,\\
				\mu_m,				&i = \cM_1, 	j = \cS, \\
				\mu_s p_r(u),		&i = \cS,		j = \cR_{\bar u}, \\
				\CC \mu_r,			&i = \cR_u, 	j = \cR_{u-1}, u \geq \CC, \\
				u \mu_r,			&i = \cR_u, 	j = \cR_{u-1}, 1\leq u < \CC, \\
				\end{cases}
\label{eq:rateBasic}
\end{equation}
%
% BIR version:
%\begin{equation}
%f({\cal p},{\cal p}^+) = \begin{cases}
				%\mu_o p_m(t),		&{\cal p} = {\cal O}, 		{\cal p}^+ = {\cal M}_{\bar t}, \\
				%\CC \mu_m,		&{\cal p} = {\cal M}_t, 	{\cal p}^+ = {\cal M}_{t-1}, t \geq \CC,  	\\
				%t\mu_m,			&{\cal p} = {\cal M}_t, 	{\cal p}^+ = {\cal M}_{t-1}, 2\leq t < \CC,\\
				%\mu_m,			&{\cal p} = {\cal M}_1, 	{\cal p}^+ = {\cal S}, \\
				%\mu_s p_r(u),		&{\cal p} = {\cal S},		{\cal p}^+ = {\cal R}_{\bar u}, \\
				%\CC \mu_r,		&{\cal p} = {\cal R}_u, 	{\cal p}^+ = {\cal R}_{u-1}, u \geq \CC, \\
				%u \mu_r,		&{\cal p} = {\cal R}_u, 	{\cal p}^+ = {\cal R}_{u-1}, 1\leq u < \CC,\\
				%\end{cases}
%\label{eq:rateBasic}
%\end{equation}
% Orginal before my changes BIR
% \begin{equation}
% f(t,j) = \begin{cases}
% 				\mu_o p_m(k),		&t=O, 	j = M_{\bar k}, \\
% 				\CC \mu_m,				&t=M_k, j = M_{k-1}, k \geq \CC,  	\\
% 				k\mu_m,					&t=M_k, j = M_{k-1}, 2\leq k < \CC,\\
% 				\mu_m,					&t=M_1, j = S, \\
% 				\mu_s p_r(k),		&t=S, 	j = R_{\bar k}, \\
% 				\CC \mu_r,				&t=R_k, j = R_{k-1}, k \geq \CC, \\
% 				k \mu_r,				&t=M_k, j = M_{k-1}, 1\leq k < \CC,\\
% 				\end{cases}
% \label{eq:rateBasic}
% \end{equation}
where ${\mathcal R}_0$ denotes the end of all reduce tasks and the job completion.

In~\eqref{eq:rateBasic} the first row corresponds to a transition from the initial setup stage $\cO$ to the map stage for a job with $t$ map tasks, \emph{i.e.}, it actually starts with $\bar t$ tasks due to early drop.
The next two rows show that the maximum parallelism is $\CC$ and that tasks finish one by one until the map stage is completed.
The next transition is to the shuffle stage $\cS$, after which the job moves on to the reduce stage, where, after dropping, a total of $\bar u$ tasks must be executed if the job has $u$ reduce tasks.
Since $N_m$ and $N_r$ are the maximum number of map and reduce tasks,
the phase space is $\cP = \{\cO, \cM_{\bar N_m},\dots, \cM_{1}, \cS, \cR_{\bar N_r},\dots, \cR_{1}\}$,
%{${\cal P} = \{{\cal O}, {\cal M}_{\bar t},\dots, {\cal M}_{1}, {\cal S}, {\cal R}_{\bar u},\dots, {\cal R}_{1}\}$},
and we can build a transition matrix $\bF$ with entries $f(i,j)$ %{$f({\cal p},{\cal Mp}^+)$}
in~\eqref{eq:rateBasic} for $i,j\in\cP$. %{${\cal p},{\cal p}^+\in\cal P$}.
Further, we define the vector $\phi = [1\quad \0]$ as the initial phase distribution, where 1 indicates that all jobs start processing in phase $i=\cO$.
The pair $(\phi,\bF)$ is thus a PH representation~\cite{latouche1999introduction} of the job processing time with ${N_m+ N_r+2}$ %\rb{}{$\bar t + \bar u + 2$} 
phases.

%The mean \emph{job} service time is $\tau = \bs{\alpha}(-\bG)^{-1}\one$, where $\one$ is a column vector of ones.
\subsection{Wave-level Model}
\label{sc:wave}
Whereas the just described model is very detailed in considering the evolution at the task level, it assumes that task execution times follow an exponential distribution. 
Generalizing this assumption at the task level is very challenging as it would require keeping track of individual tasks separately.
We thus take a different approach.
We observe that tasks tend to have fairly similar execution times~\cite{Morris:Eurosys18:sprinting}, leading to an execution in waves. 
For instance, a job composed of 40 tasks executing in a cluster with 20 computing slots will start with a first wave of 20 tasks executing in parallel. 
If these tasks have fairly similar execution times, they will finish close to each other, allowing the next 20 tasks to execute almost at the same time, making up a second wave.
This wave-level model captures this behavior, having the job processing time as a sequence of waves, each with a wave execution time.

%\RB{I'm changign all map and reduce task to t and u, no highlighting}
Given $\CC$ computing slots and a job made up of $t$ and $u$ map and reduce tasks, respectively, its effective number of map and reduce waves are
$\bar w_m = \lceil \bar t/\CC \rceil$ and $\bar w_r = \lceil \bar u/\CC \rceil$, respectively.
Recall that $\bar t = \lceil t(1-\theta_m)\rceil$ is the effective number of map tasks to execute once a task drop ratio $\theta_m$ is applied.
%%\lc{Juan, I replace i by d.}
Since waves are consecutive, we can model the execution time of the $d$-th map wave as a PH distribution with $v_{m(d)}$ phases and parameters $(\alpha_{m(d)},A_{m(d)})$, avoiding the exponential assumption and allowing for fairly general behaviors.
Moreover, we allow each wave to have a potentially different execution time, as we also observed in our experiments with state-of-the-art execution engines, \emph{e.g.}, Spark.
We similarly let the $d$-th reduce wave have a PH distribution with $v_{r(d)}$ phases and parameters $(\alpha_{r(d)},A_{r(d)})$.
Also, let the initial setup stage have $v_{o}$ phases and parameters $(\alpha_o,A_o)$,
and the intermediate shuffle stage have $v_{s}$ phases and parameters $(\alpha_s,A_s)$.
We further define the exit rate vector $a_x = -A_x\one$ for $x$ representing any of the stages considered.
Since the sum of independent PH random variables is also PH~\cite{latouche1999introduction}, we represent the job processing time as a PH distribution
with $v=v_o + \sum_{d=1}^{w_m}v_{m(d)} + v_s + \sum_{d=1}^{w_r}v_{r(d)}$ phases and parameters $(\alpha, A)$.

For clarity, consider the case where $w_m=w_r=2$, \emph{i.e.}, both map and reduce stages are composed of 2 waves of execution. 
The transition matrix $A$ of the job processing time for this case is:
%%, transition matrix given by
\begin{equation*}
\resizebox{1.01\hsize}{!}{$%
%A =
\begin{bmatrix}
A_o	&a_o\alpha_{m(1)}q_m(2)	&a_o\alpha_{m(2)}q_m(1)	&									& 											&\\
 		&A_{m(1)}								&a_{m(1)}\alpha_{m(2)}	&									&												&\\
 		& 											&A_{m(2)}								&a_{m(2)}\alpha_s	&												&	\\
 		& 											& 											&A_s							&a_s\alpha_{r(1)}q_r(2)	&a_s\alpha_{r(2)}q_r(1)		\\
 		& 											& 											& 								&A_{r(1)}								&a_{r(1)}\alpha_{r(2)}	\\
 		& 											& 											&  								& 											&A_{r(2)}								\\
\end{bmatrix}
$}%
%\label{eq:}
\end{equation*}
where $q_m(d)$ and $q_r(d)$ are the probabilities that a job requires $d$ waves of execution in the map and reduce stages, respectively.
These can be computed as
 \[
 q_m(d) = \sum_{\bar t=(d-1)C+1}^{dC} \ \sum_{t:\lceil t(1-\theta)\rceil = \bar t} p_m(t),
 \]
where the inner sum accounts for the probability that a job has $\bar t$ effective map tasks after dropping, and the outer sum accounts for all cases where the $\bar t$ effective tasks can be executed in $d$ waves.
Finally, the initial probability vector can be written simply as
$
\alpha = \begin{bmatrix} \alpha_o &0 \end{bmatrix},
$
since all jobs start in the setup stage, completing the PH representation of the job processing time at the wave level.

\subsection{Validation}

\begin{figure}[t]
	\centering
	\includegraphics[width=.66\columnwidth]{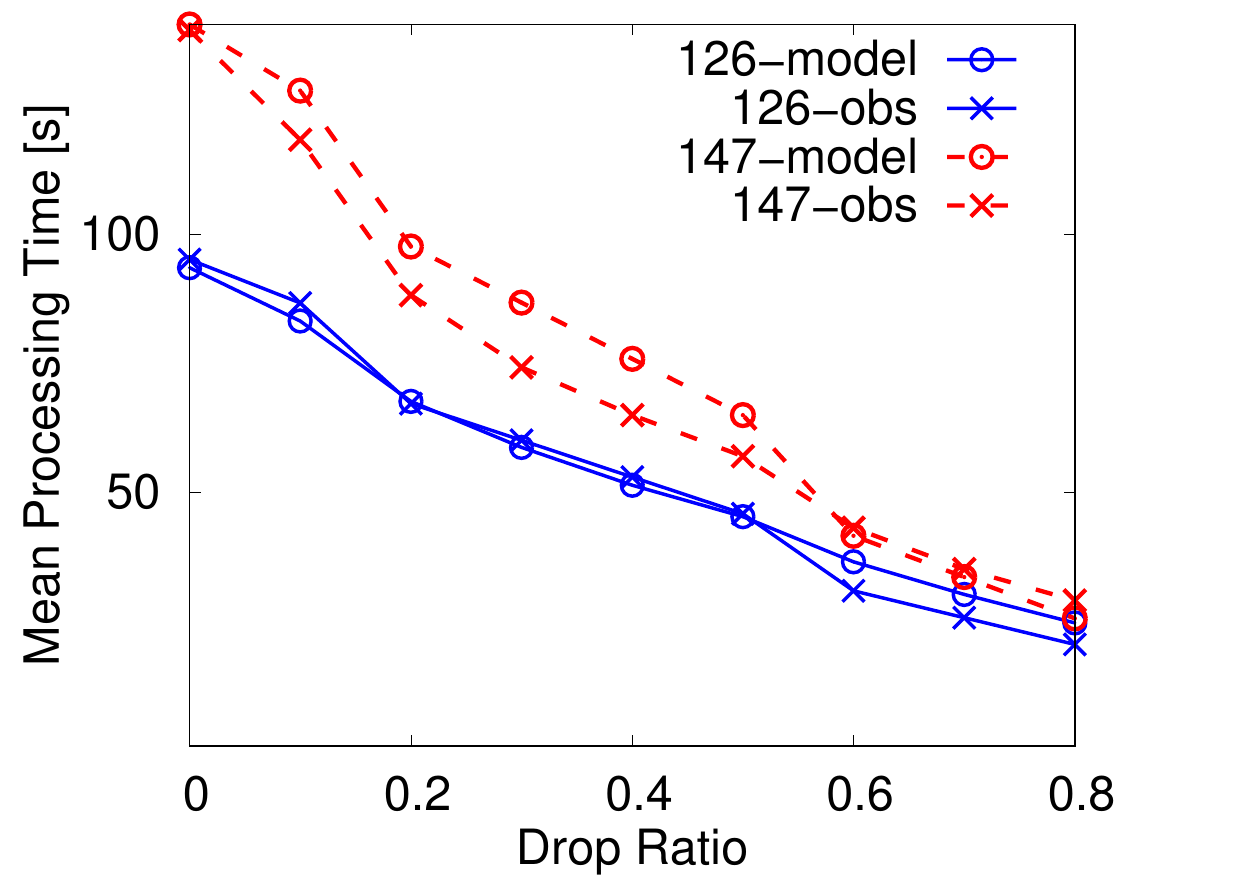}
	\caption{Validation of job processing times for different datasets and priorities.}
	\label{fig:validation_jt}
\end{figure}

We now illustrate the results obtained with the model against those observed experimentally. We first consider two different datasets, for which we obtain map and reduce task execution times from a profiling run.
The details of experiments can be found in Section~\ref{sec:evalution}.
We also collect samples of the overhead times, which we have observed to be dependent on the data size. To keep profiling at a minimum, we collect overhead times from two configurations only: one where no task drop is performed, and one where 90\% of the tasks are dropped, which is the maximum drop ratio we consider. Then, for a given drop ratio we determine the associated mean overhead time by a simple linear interpolation between these two extreme scenarios.
Figure~\ref{fig:validation_jt} shows the \emph{observed} job execution times (\emph{x} marks) for several drop ratios and two different datasets. 
It also shows the \emph{predicted} job execution times (\emph{o} marks) obtained with the model from estimations of the task execution time, the overhead, and setting the task drop ratio.
The results show that the model accurately predict the job processing time as a function of the drop ratio, with mean errors of 11.1\% and 7.8\% for the two datasets shown. 
Similar results hold for other datasets but we omit them in the interest of space.

\begin{figure}[t]
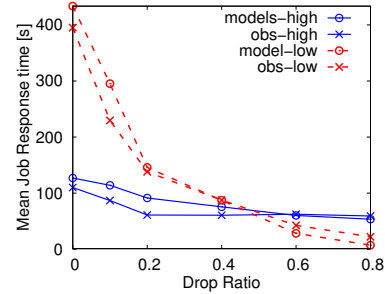

	\centering
	\includegraphics[width=.66\columnwidth]{{{figures/exp_RT_valid_exp7_ph10_util80_meanRT}}}
	\caption{Validation of response times for different datasets and priorities.}
	\label{fig:validation_rt}
\end{figure}

We now employ the model to predict the job \emph{response time}, parameterizing the model with the same information as above: mean task execution time and overhead. Also, we set the arrival rate to achieve an 80\% cluster utilization and test several values for the drop ratio. Note that for low loads the response times are similar to the processing times, which we have shown above to be accurately predicted by the model. We are thus interested in testing a high load scenario where the model must be able to predict well both the processing and waiting times.
Further, we let high- and low-priority jobs process different datasets, such that the average low-priority job size is 2.36$\times$ larger (1117MB and 473MB, respectively), and the ratio between low- and high-priority jobs is set to 9 (i.e., more low-priority jobs). This setup is similar to the ones used in the experimental section.
Figure~\ref{fig:validation_rt} shows the observed and predicted mean response times for both low- and high-priority jobs. The model is clearly able to follow the decrease in response times as the drop ratio increases, with an average error of 18.7\%.

We can therefore employ the model to determine whether a certain configuration under a given workload can achieve a preset latency objective. In fact, the model predictions can be used to determine a minimum value for %lower bound on
the drop ratio, such that the latency degradation on the high-priority jobs is kept limited. Together with a constraint on the accuracy error, it is possible to provide the user with latency-accuracy pairs for feasible drop ratios, each of which presents a different tradeoff.
\section{Evaluation}\label{sec:evalution}

This section presents our extensive evaluation of the \DA prototype atop Spark engine.
We compare it against priority systems, being preemptive or non-preemptive without approximation and sprinting.
The specific question we answer in this section is: \emph{given the accuracy requirement of multi-priority jobs, how much improvement can be obtained on the average/tail response time of low priority jobs without any resource waste and degradation of high priority jobs?}
We first describe our experimental setup, the configuration used for Spark and the workload details. 
Specifically, we focus on big data applications of text and graph analytics. %\old{in the scenario of two and three priorities}.
The number of priorities is defined based on the characteristics of Google trace which has 12 priorities but is is dominated by two to three classes that account for 89\% of all tasks ~\cite{Cavdar:greenmetrics14:priority}.
Therefore, although our proposed methodology can easily be extended to larger number of priorities, we will focus on the scenario of two and three priorities.
We first evaluate the design of differential approximation (\S\ref{sec:DA_text}), followed by the full fledged design of \DA \-- combining differential approximation and sprinting. 
While the differential approximation improves the low priority jobs at a marginal degradation of the high priority jobs, the complete \DA (\S\ref{sec:sprinting}) can improve the performance of both priorities, compared to standard preemptive and non-preemptive systems. 

%Then, we show a detailed sensitivity analysis of jobs classified into two priority levels, and a case of three priorities.
%Finally, we show the result of applying \DA on triangle count, which is a graph analysis requiring multiple map and reduce stages.

\subsection{Experimental Setup and Workloads}
\label{sec:expsetup}

\textbf{Spark processing engine.}
We rely on Spark v2.1 and a cluster with one master and ten workers. 
Each worker uses 2 CPU cores, and 4 GB memory.
Our machines consist of Dell PowerEdge R330 servers equipped with Intel Xeon E3-1270 v6 CPU, 64 hyper-threaded cores and 128 GB memory, interconnected by a 10G Ethernet switched network on a star topology.
To store the data, we deploy HDFS (v2.8.0), using one \texttt{namenode} and three \texttt{datanodes}~\cite{hdfs_conf}.

\begin{figure}[t]
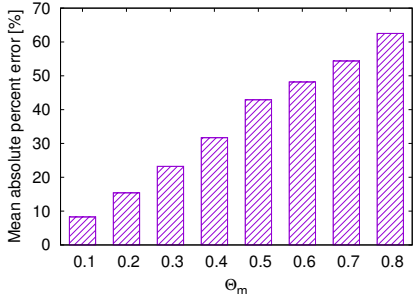

	\centering
	%\begin{tabular}{c}
		\includegraphics[width=.66\columnwidth]{{{figures/accuracy}}}\\
	%\end{tabular}
	\caption{Impact of task dropping on accuracy loss: the trend of mean absolute error.}
	\label{fig:accuracy}
\end{figure}

\textbf{Text analysis jobs.}
We deploy jobs that perform text analysis on XML data dumps collected from 164 StackEchange websites~\cite{stackexchange} each dedicated to a different topic. 
The goal of the analysis is to find the popularity of different words in different topics by first parsing the XML to extract the posts of users followed by counting the frequency of words.
Each Spark job processes one such data set. %using a DAG of RDDs. Spark runs only one concurrent task per RDD partition.
To take advantage of the 20 cores in our Spark cluster, following Spark's tuning suggestions~\cite{Spark}, we split each dataset into 50 RDD partitions.
Accordingly, Spark processes RDDs in multiple waves.
Jobs arrive following an exponentially-distributed inter-arrival time and enqueued before being dispatched to the Spark engine.
We tune the arrival rate to obtain a 80\% (50\%) system utilization based on offline profiling of our scenario.
%We set the number of partitions for Spark to 50,} which is 2.5X the total number of cores in our Spark cluster.
% \rb{Based on offline profiling, the arrival rate of XX (YY) jobs per second can achieve a system load of 80 (50)\% on our cluster.
%We modulate the job size is determined by the number of RDD partitions storing the contents of the website for analysis.
The job response time is thus composed of the queueing time and processing time.
The main metrics of interests include the average and tail response time, \emph{i.e.}, $95^{th}$, for each priority.

\textbf{Graph analysis jobs.}
We run the triangle count algorithm implemented by the Spark's \texttt{graphx}~\cite{graphx} library. 
The input dataset consists on the public Google web graph~\cite{leskovec2009community}, with 875'713 nodes and 5'105'039 edges. 
There are three types of jobs: \emph{(1)} to build the edge RDD, \emph{(2)} to build the vertex RDD, and \emph{(3)} for the triangle count itself, composed by six ShuffleMap stages and one Result stage.

\textbf{Differential approximation.}
We specifically consider scenarios of two and three levels of job priorities, with different characteristics, \emph{i.e.}, job sizes, arrival ratios across priorities, and overall system load.
As for the accuracy loss, we compute the relative errors offline under different task dropping ratios as shown in Figure~\ref{fig:accuracy}.
The mean absolute error in percentage increases sub-linearly with dropping ratios.
When dropping 10\% or 20\% of map tasks, the relative errors are roughly 8.5\% and 15\%, respectively.
Therefore, in the remainder of our evaluation section, we set the acceptable relative error to 0 for high-priority jobs and to 8.5\%, 15\% and 32\% for lower-priority jobs. 
This corresponds to evaluating the latency impact of \DA that drops 10\%, 20\% or 40\% of tasks in lower-priority jobs.
%We note that the accuracy loss caused by task dropping is independent of priority in our particular setup because the underlying data inputs are the same for both priorities.
%\RB{The last phrase goes against having different sizes for different priorities.}

\begin{figure}[t]
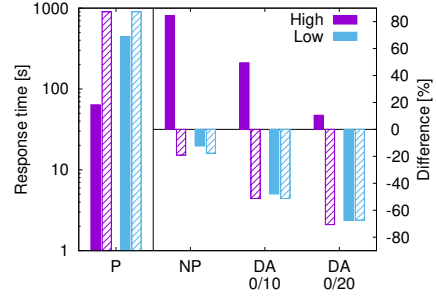

  \centering
  %\begin{tabular}{c}
    \includegraphics[width=.66\columnwidth]{{{figures/reference-exp7-rel}}}\\
   %\end{tabular}
  \caption{Mean (solid bars) and tail (shaded bars) latency improvement of two differential approximation variations on a two-priority system. $P$: preemptive, absolute values. $NP$: non-preemptive. $NP$, $DA_{(0,100)}$, and $DA_{(0,20)}$ as relative difference to $P$.}
  \label{fig:reference}
\end{figure}

\begin{figure*}[t]
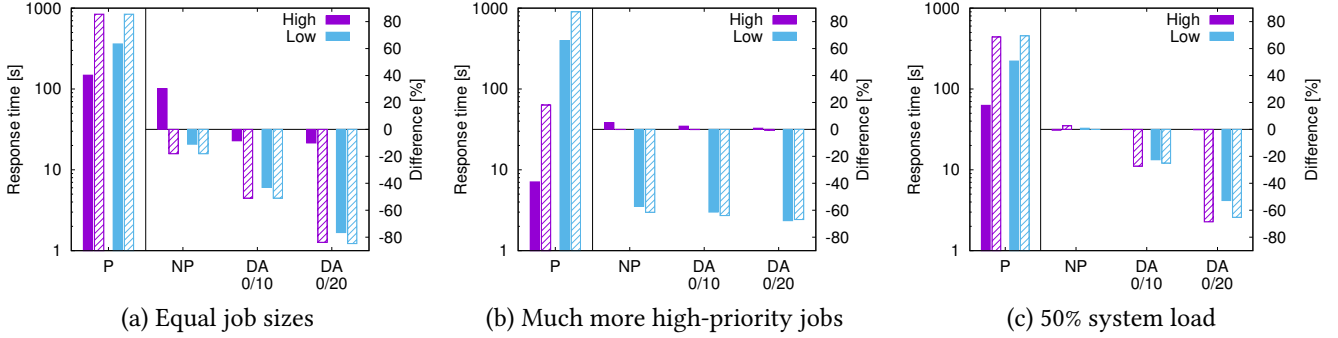

	\centering
	\begin{tabular}{ccc}
		\includegraphics[width=.66\columnwidth]{{{figures/lower-dataset-ratio-rel}}}&
		\includegraphics[width=.66\columnwidth]{{{figures/opposite-weight-ratio-rel}}}&
		\includegraphics[width=.66\columnwidth]{{{figures/lower-load-rel}}}\\
		(a) Equal job sizes &  (b) Much more high-priority jobs & (c) 50\% system load
	\end{tabular}
	\caption{Sensitivity analysis of differential approximation on mean (solid bars) and tail (shaded bars) latencies by changing the ratios between high and low priority jobs: sizes, arrival rates, and the overall system utilization.}
	\label{fig:prior2-Sensitivity}
\end{figure*}

\textbf{Differential sprinting.}
%\RB{Please check numbers highlighted}
We use DVFS as the sprinting mechanism to change the speed of the CPU. 
The CPU clock frequency is initially set to 800MHz old and temporarily increase it to 2.4GHz old when
sprinting. %800MHz and temporarily increase it to 2.4GHz when sprinting.
These frequencies were defined based on the limits supported by the machines used, which also corresponds to a common setup \cite{frequency}.
Sprinting reduces the execution time of high priority jobs by up to 60\%, but increases the servers power consumption by 1.5x, from 180W to 270W. 
In the following, we consider two types of energy budgets:
\emph{(1)} limited sprinting, to sprint only 35\% of the execution time of high priority jobs, and \emph{(2)} unlimited sprinting, to sprint high priority jobs for their whole duration. 
Under limited sprinting high priority jobs sprint after 65 seconds. 
Under unlimited sprinting they sprint as soon as they are dispatched.

\textbf{Resource waste.}
With \DA, differential approximation and sprinting levels are applied on different priorities and lower-priority jobs are never evicted upon arrival of a higher-priority job. 
As a result, machine time is never wasted on reprocessing evicted jobs compared to a preemptive priority system.
We define the resource waste as the percentage of machine time used to re-process evicted jobs compared to the total processing time.
%\RB{WE have this data (column  but currently it is not shown. Do we add it somewhere or we remove the last phrase?}
\subsection{Differential Approximation} \label{sec:DA_text}
\subsubsection{Two-Priority System}

We first show the effectiveness of differential approximation on a reference setup, highlighting the difference of mean and $95^{th}$ latency for both high- and low-priority jobs when compared to a preemptive and a non-preemptive priority system denoted as $P$ and $NP$, respectively.
The three key parameters in the reference setup are: \emph{(i)} the ratio between low- and high-priority jobs is 9 to 1, \emph{(ii)} the average sizes of low- and high-priority jobs are 1117MB and 473MB, respectively, \emph{(iii)} and the average system load is 80\%. The parameters are set as close as to the workload characteristics of  Google trace~\cite{Clusterdata:Wilkes2011}.
Figure~\ref{fig:reference} summarizes the absolute results of the preemptive priority setup, and its relative difference compared to a non-preemptive priority setup, $DA_{(0,10)}$ and $DA_{(0,20)}$. 
The subscript pair of $DA$ denotes the task dropping ratio for high- and low-priority respectively. 
We use solid bars for the mean latency, while shaded bars are for the $95^{th}$ percentile latency. The resource waste is roughly 4\% with the preemptive priority policy.

Under $P$, the mean latency of high-priority is better than the low-priority job. This stems from the unbalance in the queueing times: 0.03s versus 310s on average for high- and low-priority jobs, respectively. 
This difference is smaller for the $95^{th}$ percentile.
When using the $NP$ model, where preemption of low-priority jobs is not allowed, the performance of low-priority jobs improves roughly by 20\% at the cost of increasing the latency of high-priority jobs by 80\%. 
This is because high-priority jobs have to wait for the low-priority jobs in execution to finish before getting served. 
In contrast, $DA_{(0,20)}$ can significantly improve the performance (roughly 65\%) of both mean and tail latency of low-priority at only a marginal (10\%) increase in the mean latency of high-priority jobs and an accuracy loss for low-priority jobs of 15\%.

We further consider a use case scenario where it is possible to tolerate a 30\% accuracy loss for low-priority jobs while maintaining the latency of high-priority jobs under 100ms with no accuracy loss.
The task deflator consults the results in Figure~\ref{fig:accuracy} to determine the maximum drop ratios to attain an accuracy target of 0\% and 30\% for high- and low-priority jobs, respectively.
Likewise, the deflator runs the \DA model (Section~\ref{sec:model}) and determines that a 20\% drop ratio for low-priority jobs is already within the 100ms limit for the high-priority mean latency (as Figure~\ref{fig:validation_rt} shows). 
We can thus choose to employ $DA_{(0,20)}$ to hold both accuracy and latency constraints, as confirmed by the experimental results in Figure~\ref{fig:reference}. 
This selection can be easily automated by assigning weights to the latency and accuracy targets to select among the feasible drop ratios.

\subsubsection{Sensitivity Analysis}\label{subsec:sensitivity}

Our sensitivity analysis of differential approximation fiddles with following parameters in the reference setup one at a time: \emph{(i)} high- and low-priority jobs of same size, \emph{(ii)} ratio between low- and high-priority jobs set 1 to 9, and \emph{(iii)} a total arrival rate resulting in a 50\% system load.
Figure~\ref{fig:prior2-Sensitivity} summarizes the results for these three scenarios.
Due to the rich information embedded in the figure%and the space limit
, we focus on comparing the latency gains of differential approximation between the reference and new setup.

\textbf{Similar job size for both priorities}.
Comparing Figure~\ref{fig:prior2-Sensitivity} (a) and Figure~\ref{fig:reference}, the latency gain of differential approximation is significant, \emph{i.e.}, for low-priority up to 80\%.
%Moreover, there is a significant improvement for 
High-priority jobs improves too: both their mean and tail latencies have better improvement than the reference system.
This can be explained by the fact that high-priority jobs have shorter waiting time here than in the reference system. 
In a non-preemptive setting, being in $NP$ or $DA$, the maximum amount of queueing time for an arriving high-priority job is a single execution of low-priority job, assuming the high-priority queue is empty. 
Hence, smaller the low-priority jobs, better the gain of differential approximation used in the non-preemptive setup.

\textbf{Relatively increased high- to low-priority job ratio}.
Comparing Figure~\ref{fig:prior2-Sensitivity} (b) and Figure~\ref{fig:reference}, the latency gain of differential approximation is worse.
Both the mean and tail latency of high-priority increase considerably.
Though the average latency gain of low-priority remains the same as the reference case, the tail latency gain decreases from 60\% to 20\%.
As differential approximation only applies approximation techniques on low-priority jobs which account for 10\% of the total jobs, its effectiveness is limited.
Hence, in the scenario of dominant high-priority jobs, one shall activate approximation for both priorities.
%\RB{The last phrase, might open up the question why did you not run it?}

\textbf{Relatively low system loads}.
Comparing Figure~\ref{fig:prior2-Sensitivity} (c) and Figure~\ref{fig:reference}, the latency gain of $DA_{(0,10)}$ is slightly worse, but $DA_{(0,20)}$ maintains a similar gain as the reference setup.
%Another noteworthy observation is that 
Further, there is almost no performance degradation from preemptive to non-preemptive system, shown by the results of $NP$.
When the system load is low, \emph{e.g.}, 50\%, there is no difference between preemptive and non-preemptive priority systems because the engine is rarely occupied when higher-priority jobs arrive.
The gain of $DA_{(0,20)}$ on low-priority jobs is thus mainly attributed to the reduction of processing time, instead of queueing time.
Moreover, the difference between $DA_{(0,10)}$ and $DA_{(0,20)}$ can be explained by the fact that dropping 20\% of tasks reaches the critical mass to drop an entire wave.
Overall, thanks to flexible approximation levels and stochastic models of deflator, differential approximation can effectively tradeoff analysis accuracy for improved mean/tail latencies of high/low-priority jobs, against complex system and workload dynamics.
\subsubsection{Three-Priority System}

\begin{figure}[t]
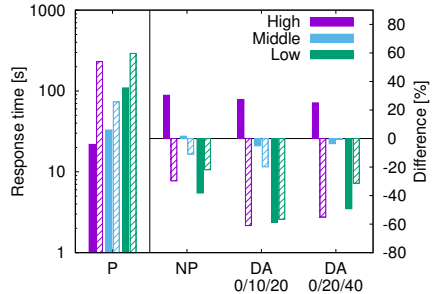

  \centering
  %\begin{tabular}{c}
    \includegraphics[width=.66\columnwidth]{{{figures/3-priority-classes-rel}}}
   %  \end{tabular}
  \caption{Differential approximation on three-priority system: relative difference in mean (solid bars) and tail (shaded bars) latency against preemptive priority ($P$).}
  \label{fig:3prior}
\end{figure}

We demonstrate the performance gains for differential approximation on a system with three priorities: high, medium and low (Figure~\ref{fig:3prior}).
The total arrival rate is 2.3 jobs/min with rate ratio of high-medium-low priority of 1-4-5, resulting in roughly 80\% system load.
For drop rates, we use $DA_{(0,10,20)}$ and $DA_{(0,20,40)}$: the former introduces~8.5(15)\% accuracy loss for medium(low)-priority jobs, and the latter introduces~15(32)\% accuracy loss for medium(low)-priority jobs according to Figure~\ref{fig:accuracy}.

Similar to the two-priority scenario, we use the mean/tail latency of the preemptive priority setup as comparison baseline.
The resource waste under $P$ is roughly 16\%. 
The remaining three setups in Figure~\ref{fig:3prior} incur in zero resource waste, due to their non-preemptive nature.
In terms of latency improvement, differential approximation is able to significantly reduce the tail latency for all three priorities by up to 60\%.
Differential approximation reduces the average latency more for low-priority than medium-priority.
However, such improvement of differential approximation comes at the cost of slightly higher average latency of high-priority jobs and accuracy loss of low- and medium-priority jobs.
In this particular setup, $DA_{(0,10,20)}$ appears to achieve the most moderate tradeoff among accuracy/latency for high/lower-priority~jobs.
%Our model of deflator also further suggests that one shall consider applying approximation level on the high priority jobs so as to improve its average latency.
%\RB{The last phrase, might open up the question why did you not run it?}

\begin{figure}[t]
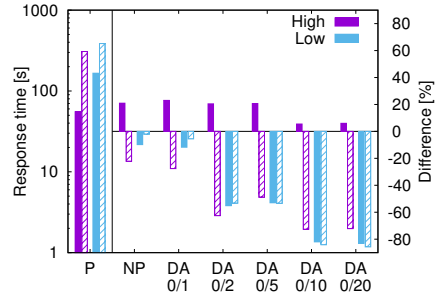

	\centering
        %\begin{tabular}{ccc}
		\includegraphics[width=.66\columnwidth]{{{figures/reference-trianglecount-trianglecount-exp7-rel}}}
		%\end{tabular}
	\caption{Differential approximation on triangle count: relative difference in mean (solid bars) and tail (shaded bars) latency against preemptive priority scheduler ($P$).}
	\label{fig:3priorTria}
\end{figure}

\begin{figure*}[t]
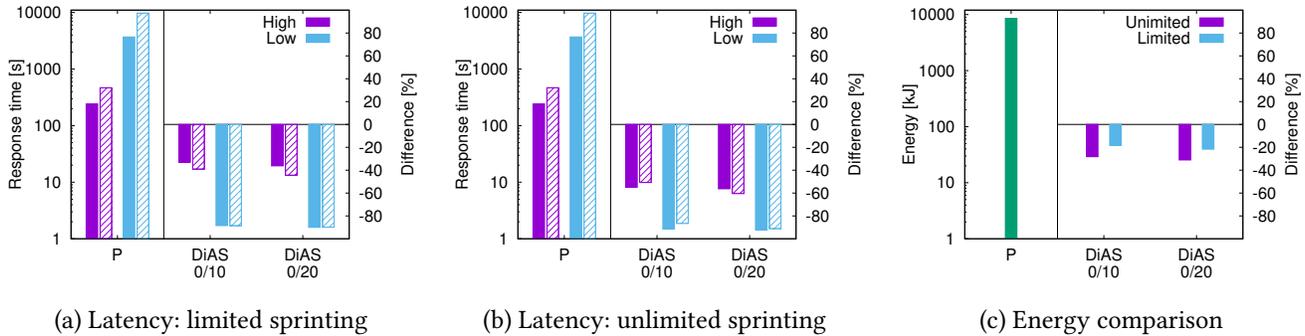

	\centering
	\begin{tabular}{ccc}
		\includegraphics[width=.66\columnwidth]{{{figures/exp-limited-sprinting}}} 
		& \includegraphics[width=.66\columnwidth]{{{figures/exp-unlimited-sprinting}}}
		&\includegraphics[width=.66\columnwidth]{{{figures/energy}}}\\
		\vspace{-5pt}
		(a) Latency: limited sprinting & (b) Latency: unlimited sprinting & (c) Energy comparison\\
	\end{tabular}
	\vspace{-5pt}
	\caption{Complete \DA on triangle count:  latency and energy improvement against the preemptive priority scheduler (P). In (a) and (b), the mean and tail latency are in solid bars and shaded bars, respectively.}
	\vspace{-5pt}
	\label{fig:sprinting}
\end{figure*}

\subsubsection{Differential Approximation on Triangle Count}
\label{sec:DA_graph}
We further illustrate the gains of differential approximation when the computation requires several map and reduce stages. 
Specifically, we run the triangle count algorithm implemented in \texttt{graphx} library in Spark. %As input we employ the graph published by Google in its web competition~\cite{leskovec2009community}. %The program is made of three jobs: one to build the edge RDD, one to build the vertex RDD, and one for the triangle count itself. The triangle count job is made of six ShuffleMap stages and one Result stage.
Task dropping in this case is performed on every ShuffleMap stage, for which we consider drop ratios \{1,2,5,10,20\} for the low-priority jobs.
%Note that since we drop in every ShuffleMap stage,
The total effective drop ratio is thus the result of applying the stage drop ratio in each stage.

Figure~\ref{fig:3priorTria} displays the gains obtained with differential approximation with respect to preemptive and non-preemptive scheduling.
Clearly, with fairly limited task dropping ratios (5-10\%) differential approximation is able to reduce the mean latency of low-priority jobs by over 50\%.
Moreover, differential approximation reduces by a similar factor the tail latency of \emph{both} high and low-priority jobs.

%The drop ratio of \{1,2,5,10,20,30,40,50,60\} applies to each MapShuffle stage. 

\subsection{Differential Approximation and Sprinting}\label{sec:sprinting}

Finally, we evaluate the complete design of \DA, applying different CPU sprinting on the high priority jobs and approximation on the low priority jobs.  %effectiveness of applying sprinting to our differential approximation approach.
We consider graph analytics jobs, which has high and low priorities of the same job size with a ratio of 3 to 7.
We experiment the CPU sprinting policy under two different energy budgets resulting in two different scenarios.
In the first scenario, \emph{i.e.}, limited sprinting shown in Figure~\ref{fig:sprinting} (a), we consider a sprinting budget of 22kJ which roughly limits the jobs to run in high frequency only for 35\% of their execution time based on timeout.
In the second one, \emph{i.e.}, unlimited sprinting shown in Figure~\ref{fig:sprinting} (b), we set the budget high enough such that the high-priority jobs run at high frequency throughout their whole execution time.
%We apply the same sprinting policies also to the non-preemption system marked as $NPS$ in the figures, 
We use a non-sprinted $P$  system as the baseline for the ease of comparison.

\textbf{Latency gain}. The complete \DA of differential approximation and sprinting shows promising performance. First, the average and tail latency of \emph{both priorities} improve under limited and unlimited sprinting budgets, ranging between 35\% to 90\%.  Overall, the latency gain is more prominent for the tail latency,  low-priority jobs,  and unlimited sprinting.
In terms of absolute comparison, the improvement for low-priority jobs is  around 90\%, whereas high-priority is between 40-60\% depending on the sprinting budget. 
We stress that few cases with increased average latency of high-priority jobs observed in Section~\ref{sec:DA_text} are effectively countered by enabling differential sprinting.
The performance gains are therefore more consistent for both tail and average latency, compared to the approximation-only results.

Despite the focus of sprinting being on high-priority jobs, the response times of low-priority jobs are also indirectly improved.
When compared to differential approximation only, the average response time of $\DA_{(0,20)}$ increases up to 55\% for high-priority jobs but also up to 40\% for low priority jobs.
Similarly, for $\DA_{(0,10)}$ the increase goes up to 50\% and 53\% for high- and low priority jobs, respectively. 
That is, by reducing the processing time of high-priority jobs, the queueing time of low-priority jobs is reduced, directly affecting the response time for both types.
%More details are shown in Figure~\ref{fig:da-vs-dias}.
%\ir{Maybe I could add a bar plot showing this comparison in more details?}\vs{yes}
%\lc{Could we explain why P results are so good?}\RB{see comment above}

%\begin{figure}[t]
%	\centering
%	\includegraphics[width=.66\columnwidth]{{{figures/energy}}}\\
%	\caption{Differential approximation on triangle count: relative difference in mean (solid bars) and tail (shaded bars) latency against preemptive priority ($P$).}
%	\label{fig:energy}
%\end{figure}

\textbf{{Latency decomposition}}. To unveil the exact performance advantage of \DA, we zoom into the performance of the limited sprinting case and present the average queueing and execution times for high- and low- priority jobs in Table~\ref{tab:time}. We also apply the same sprinting policy on the non-preemptive system, termed $NPS$. Due to sprinting, the execution times of high-priority jobs are lower than the low-priority jobs by at least 25\%. Because of the 20\% task dropping in $\DA_{(0,20)}$, the average execution time of the low-priority jobs is the lowest among the three policies, \emph{i.e.}, around $131$~seconds. The percentage of time that low-priority jobs occupy the system thus reduces, avoiding longer waiting time for both high- and low-priority jobs. As such, the queueing times for both high- and low-priorities are lower than $NPS$ and~$\DA_{(0,10)}$.

% the following table is for the unlimited case
%\begin{table}[htp]
%\caption{Average queueing and execution times of high- and low-priority jobs under sprinted non-preemptive scheduling, $\DA_{(0,10)}$ and $\DA_{(0,20)}$. }
%\label{tab:time}
%\begin{footnotesize}
%\begin{tabular}{l|l|l|l|l|l|l|}
%\cline{2-7}
%                           & \multicolumn{2}{l|}{$NPS$}    & \multicolumn{2}{l|}{$\DA_{(0,10)}$}   & \multicolumn{2}{l|}{$\DA_{(0,20)}$}    \\ \cline{2-7} 
%                           & Queue {[}s{]} & Exe.{[}s{]} & Queue {[}s{]} & Exe.{[}s{]} & Queue {[}s{]} & Exe. {[}s{]} \\ \hline
%\multicolumn{1}{|l|}{High} & 49.0          & 58.6        & 50.2          & 58.8        & \textbf{44.7 }      & 61.9         \\ \hline
%\multicolumn{1}{|l|}{Low}  & 156.3         & 148.2       & 174.3         & 139.4       & \textbf{155.4 }       &  \textbf{130.5}       \\ \hline
%\end{tabular}
%\end{footnotesize}
%\end{table}

\begin{table}[t]
\caption{Average queueing and execution times of high- and low-priority jobs under sprinted non-preemptive scheduling (\emph{NPS}), $\DA_{(0,10)}$ and $\DA_{(0,20)}$. }
\label{tab:time}
\begin{footnotesize}
\setlength{\tabcolsep}{4pt}
\rowcolors{1}{gray!10}{gray!0}
\begin{tabular}{lllllll}
	\rowcolor{gray!25}
%\cline{2-7}
                             & \multicolumn{2}{c}{\textbf{NPS}}    & \multicolumn{2}{c}{\textbf{$\DA_{(0,10)}$}}   & \multicolumn{2}{c}{\textbf{$\DA_{(0,20)}$}}    \\ \cline{2-7} 
                           & Queue {[}s{]} & Exe.{[}s{]} & Queue {[}s{]} & Exe.{[}s{]} & Queue {[}s{]} & Exe. {[}s{]} \\ \hline
\multicolumn{1}{l}{High} & 70.6          & 99.8        & 70.0          & 100.2       & \textbf{ 55.1 }         & 99.4         \\ \hline
\multicolumn{1}{l}{Low}  & 378.9         & 148.5       & 286.42        & 139.0       &\textbf{ 238.0}         &  \textbf{131.1}       \\ \hline
\end{tabular}
\vspace{-5pt}
\end{footnotesize}
\end{table}

\textbf{Energy gain}.  In Figure~\ref{fig:sprinting}(c), we summarize the normalized energy consumption of \DA against the $P$ policy. For both unlimited and limited sprinting, we temporarily increase the CPU frequency for high-priority jobs. 
One would expect a slightly higher energy consumption, compared to the no-sprinting baseline. 
Surprisingly, for both unlimited and limited cases, \DA reduces the overall energy consumption. 
The energy reductions stemming from differential sprinting alone for the limited and unlimited budgets are around 15\% and 26\%, respectively. 
We explain this result by the significant reduction in execution times that outweighs the power increase during sprinting. 

The energy gain of \DA can also be amplified by the approximation ratios, \emph{i.e.}, $\DA_{(0,10)}$ and $\DA_{(0,20)}$. 
With unlimited sprinting the gain increase to 28.2\% and 31\% for $\DA_{(0,10)}$ and $\DA_{(0,20)}$, respectively.
Similarly, for limited sprinting we observe 18.3\% and 21.6\%.
Higher reductions are observed for higher drop rates, as dropping reduces the computational load on the cluster.
%Moreover, we observe that the scenario with limited sprinting profits more from the differential approximation of low-priority jobs. Here the gains increase by up to 7 and 5 percent points for high- and low-priority jobs, respectively. \ir{commenting this about the comparison with NPS since we also removed it from the plot.}
Overall, the design of combining differential approximation and differential sprinting can improve both the latency of both priorities and energy consumptions across diversified systems scenarios.

We suggest the following procedure to determine the static threshold used in the algorithm.  
To utilize the proposed models to predict the performance, one needs to first obtain the input parameters of the proposed models through workload profiling, that quantifies the relationship between file size and execution time under a constant CPU frequency. 
Then, one can exhaustively search through different combinations of dropping ratios, priorities, and frequency thresholds. 
Our proposed models can estimate the latency of such large combinations quickly. 
The values that optimize the tradeoff are then selected for a given set of workloads. 
We note that such searching procedure needs to be evoked upon every workload changes. 
\DA considers a scenario where the workload set is given and hence only consider the static~threshold.
  
%\lc{could we provide the energy consumption comparison between limited and unlimited case}

%\begin{figure}[t]
%	\centering
%	\includegraphics[width=.66\columnwidth]{{{figures/da-vs-dias}}}
%	\caption{Unlimited sprinting (US) and limited sprinting (LS) in comparison to differential approximation. }
%	\label{fig:da-vs-dias}
%\end{figure}

\section{Related Work}\label{sec:related}
%To cater to the emergence of the big data era, 
A plethora of state-of-the-art systems developed novel priority-aware or approximation/sprinting-enabled processing platforms.
%aimed at improving the latency and computation efficiency of big data jobs.
We particularly highlight those focused on managing priority, approximation frameworks and sprinting strategies for Spark-like applications.
Moreover, we summarize efforts on computational sprinting and modelling that address the challenging question -- latency distribution for multi-priority jobs composed of multiple parallel tasks.

\textbf{Priority systems.}
Characterization studies~\cite{Chen:VLDB12:FacebookClaudera,Cavdar:greenmetrics14:priority,Rosa:DSN15:failure} on production big data systems show that multi-task jobs are associated with multiple priorities and exhibit diverse workload characteristics~\cite{Chen:VLDB12:FacebookClaudera}.
To ensure the performance of (particularly high) priority jobs, the scheduler evicts low-priority jobs to make resource available for high priority ones, resulting in a significant resource waste~\cite{Cavdar:greenmetrics14:priority} and latency penalty on the low priority jobs~\cite{Rosa:DSN15:failure}.
Indeed, modern big processing engines, \emph{e.g.}, Hadoop~\cite{hadoop} and Spark~\cite{Spark}, also support multi-priority job scheduling.
For instance, Hadoop's fair scheduler~\cite{FairScheduler} can assign different weights on different workloads to achieve soft priority, \emph{i.e.}, higher (lower) weights on higher (lower) priority.
Mesos~\cite{Hindmand:NSDI11:Mesos} is a cluster manager that support priorities across and within multiple processing engines with a focus on fairness. 
Omega~\cite{Schwarzkopf:Eurosys13:omega} is a two-level priority scheduler designed for large-scale system.
Recognizing the need of evictions in priority systems, Natjam~\cite{Cho:SoCC13:Natjam} develops novel job and task eviction policies for a scenario of two priorities that have different deadlines.
%In contrast to the related work on novel scheduling approaches of tasks and jobs, 
\DA proposes an orthogonal solution that alters the jobs resource demands and processing speeds for different priorities, rather than a novel scheduling approach of tasks or jobs.

\textbf{Approximation big data engines.}
To process vast and fast amount of data influx, novel approximation-enabled systems are designed to meet the dual objectives of accurate analysis and resource efficiency.
In the context of MapReduce paradigm, statistical sampling theory is commonly applied to selectively process a subset of data either at the level of input block~\cite{Agarwal:Eurosys:BlinkDB} or task~\cite{Krishnan:WWW16:IncApprox}, before or after the execution starts.
BlinkDB \cite{Agarwal:Eurosys:BlinkDB}, an approximate query processing framework, provides accuracy guards in short response times by leveraging statistical sampling theory to choose the inputs.
ApproxHadoop~\cite{Goiri:ASPLOS15:APPROXHADOOP} develops a two-stage sampling strategy for Hadoop~\cite{hadoop} by either dropping the tasks or amount of data per tasks, so as to minimize the overhead of data accessing.
To overcome the accuracy loss of sampling, IncApprox~\cite{Krishnan:WWW16:IncApprox} combines the task sampling and incremental computing, \emph{i.e.}, memorizing intermediate historical result.
Grass~\cite{Ganesh:NSDI14:GRASS} is a scheduler that prioritizes jobs with higher approximation level over lower levels for approximate analytics engine.

While approximate big data engines effectively trade accuracy for the latency target and resource efficiency, they only consider single-priority scenarios and often overlook the latency models of complex dependency of jobs arrival.

\textbf{Computational sprinting.}
Computational sprinting enables bursts of peak processing in systems limited by dark silicon. 
Sprinting mechanisms exist nowadays at all system levels, from transistors and processors~\cite{DBLP:journals/micro/RotemNAWR12}, to computer systems~\cite{DBLP:conf/iccd/KomodaHNMN13}, %virtual machines~\cite{DBLP:conf/eurosys/WangUGKL17}, 
and datacenters~\cite{DBLP:conf/icdcs/ZhengW15}. %and cooling systems~\cite{DBLP:conf/asplos/GoiriNB15}.  
Sprinting policies decide when and what to sprint, \emph{e.g.}, phases within job executions \cite{zhang2016pupil} and particular queries ~\cite{DBLP:conf/asplos/JeonHKERC16}.  
Several approaches have been explored in the single priority scenario,  from simple heuristics~\cite{DBLP:conf/hpca/HsuZLMWMTD15}, to queueing~\cite{DBLP:conf/infocom/QiuPH16} and machine learning~\cite{Morris:Eurosys18:sprinting} models.
\DA extends the sprinting policies with multi-priority scheduling and approximation schemes such that the performance of all priorities can be improved.

\textbf{Stochastic models for multi-priority jobs.}
Modeling the latency for multi-priority jobs is a long standing challenge by itself because of the complex workload dynamics across jobs and the interdependency among tasks.
For priority systems, modeling studies can be categorized into single \emph{vs.} multi-server setus, preemptive \emph{vs.} non-preemptive, and resume \emph{vs.} non-resume under preemptive scheduling.
 ~\cite{Mor:Questa05:priority, Wierman:2006:Performance} employ matrix analytics method to analyze the jobs average latency in the non-preemptive multi-server system, whereas \cite{sleptchenko2005} focus on the state probabilities of preemptive multi-server systems.
Horv{\'{a}}th~\cite{Horvath:ejor15:priority} derives the latency distribution in both preemptive and non-preemptive setting for single server system.
Jelenkovic~\cite{Jelenkovic2014} derives the stability conditions for non-resume preemptive systems, highlighting the high risk of instability.

%To the best of our knowledge, no existing system tackles the latency modeling challenge of multi-task, multi-priority jobs.
The stochastic model in \DA not only captures the entire distribution of latency for multi-priority and multi-task jobs but also further facilities the optimization of differential approximation and sprinting for real system deployment.

\section{Concluding Remarks}
%Motivated by the daunting resource waste and severe latency degradation of low-priority jobs in production systems,
We propose a novel design of differential approximation and sprinting, \DA, to trades off the accuracy and additional sprinting capacity for improving the efficiency of big data engines, \emph{i.e.}, reduction of mean/tail latency without resource waste.
The design of \DA is general, \emph{i.e.}, supports different types of analyses and multiple priorities, and compatible with existing MapReduce based processing engines that provide approximation mechanisms, \emph{e.g.}, task dropping, and dynamic frequency scaling.
%We derive stochastic analysis using matrix analytic method to guide \DA to answer the following key question: how many tasks of lower priority jobs to drop so as to fulfil the accuracy and mean/tail latency requirements across all priorities?
We derive stochastic models to guide the control of approximation and sprinting levels of \DA. 
We implement the prototype of \DA atop of Spark, with examples of text and graph analytics.
% and extensively evaluate scenarios encompassing different number of priorities, their relative ratio, job size ratio, and load intensities.
Our extensive evaluation results show that %that \DA effectively strike the tradeoff between analysis latency, accuracy without any resource waste.
\DA consistently reduces the mean/tail latency of both low- and high-priority jobs (by up to 90\% and 60\%, respectively) at roughly 15\% relative error in the accuracy of low-priority jobs and more than 20\% energy reduction, compared to the state-of-art  preemptive scheduler.  

\section*{Acknowledgment}
%This work has been partly funded by SNSF projects  407540\_167266 and 200021\_141002.
%The research of Juan F. P\'erez has been supported by the ARC Centre of Excellence for Mathematical and Statistical Frontiers (ACEMS).
The research leading to these results has received funding from the European Union's Horizon 2020 research and innovation programme under the LEGaTO Project (\url{legato-project.eu}), grant agreement No~780681. 
This work has been partly funded by the Swiss National Science Foundation NRP75 project 407540\_167266.
\bibliographystyle{ACM-Reference-Format}
\bibliography{priority,StreamSpark,other}

\balance

\clearpage
\end{document}